\documentclass[prd,aps,twocolumn,nofootinbib,preprintnumbers,superscriptaddress,balancelastpage,longbibliography]{revtex4-2}

\usepackage{graphicx}
\usepackage{hyperref}
\usepackage{aas_macros}
\usepackage{amsmath}
\usepackage[dvipsnames]{xcolor}
\usepackage[normalem]{ulem}

\hypersetup{
     colorlinks   = true,
     citecolor    = teal,
     urlcolor     = teal,
     linkcolor    = teal
}

\begin{document}

\title{Solar Spin-Dependent Dark Matter-Neutron Cross Section Constraints:\\ The Lost Case}

\author{Thong T.Q. Nguyen}
\thanks{{\scriptsize Email}: \href{mailto:thong.nguyen@fysik.su.se}{thong.nguyen@fysik.su.se}; \href{https://orcid.org/0000-0002-8460-0219}{0000-0002-8460-0219}}
\affiliation{Stockholm University and The Oskar Klein Centre for Cosmoparticle Physics, Alba Nova, 10691 Stockholm, Sweden}

\author{Carlos Blanco}
\thanks{{\scriptsize Email}: \href{mailto:carlosblanco2718@princeton.edu}{carlosblanco2718@princeton.edu}; \href{https://orcid.org/0000-0001-8971-834X}{0000-0001-8971-834X}}
\affiliation{Institute for Gravitation and the Cosmos, The Pennsylvania State University, University Park, PA 16802, USA}
\affiliation{Department of Physics, Princeton University, Princeton, NJ 08544, USA}
\affiliation{Stockholm University and The Oskar Klein Centre for Cosmoparticle Physics, Alba Nova, 10691 Stockholm, Sweden}

\author{Tim Linden}
\thanks{{\scriptsize Email}: \href{mailto:linden@fysik.su.se}{linden@fysik.su.se}; \href{https://orcid.org/0000-0001-9888-0971}{0000-0001-9888-0971}}
\affiliation{Stockholm University and The Oskar Klein Centre for Cosmoparticle Physics, Alba Nova, 10691 Stockholm, Sweden}

\begin{abstract}
\noindent The Sun has long served as a natural dark matter detector, capturing halo particles that scatter with solar nuclei and electrons and subsequently produce indirect signals from dark matter annihilation that can be probed by either neutrino or $\gamma$-ray instruments. Previous efforts have computed cross-section constraints for dark matter scattering with electrons, as well as spin-independent and spin-dependent dark matter-proton scattering. However, the spin-dependent dark matter-neutron scattering scenario has been neglected, despite its importance in direct detection. For the first time, we compute the capture and evaporation rates for spin-dependent dark matter-neutron scattering in the Sun, including odd-neutron isotope targets in the current standard solar model. Comparing the predicted annihilation signals against current neutrino and $\gamma$-ray observations, and projecting the reach of upcoming detectors, we find that solar constraints exceed direct detection limits for several annihilation channels. For models in which the dark matter annihilates into long-lived mediators, these constraints can extend below the neutrino fog.
 \end{abstract}

\maketitle

\section{Introduction}
\label{sect:intro}

Identifying the non-gravitational interactions between dark matter (DM) and the Standard Model (SM) remains one of the central missions of modern physics~\cite{Bertone:2016nfn, Bertone:2004pz, Bertone:2018krk, Cirelli:2024ssz}. Alongside the terrestrial direct-detection program~\cite{Baudis:2025yva, Akerib:2022ort, Essig:2022dfa, Chou:2022luk}, a complementary strategy proposed over forty years ago uses the Sun itself as a natural detector~\cite{Press:1985ug, Griest:1986yu, Srednicki:1986vj}. In this scenario, halo DM particles scatter with solar nuclei and electrons, become gravitationally bound, and accumulate over the age of the Sun. Trapped DM particles can annihilate into SM final states, leaving indirect-detection observables such as neutrinos and $\gamma$-rays that can be detected by existing telescopes.

Since then, the calculations of DM capture, annihilation, and evaporation in the Sun have been developed rigorously~\cite{Gould:1987ju, Gould:1989hm, Gould:1989tu, Garani:2017jcj, Garani:2021feo, Banks:2021sba, Busoni:2017mhe, Vincent:2015gqa, Vincent:2014jia, Busoni:2013kaa, Acevedo:2023owd, Leane:2022hkk, Bramante:2017xlb, Leane:2023woh}, establishing a mature framework for predicting the DM-induced signal~\cite{Leane:2024bvh, Acevedo:2020gro, Liu:2020ckq, Nguyen:2026nhe}. Furthermore, both theorists and experimental collaborations have used the solar DM capture scenario to investigate both spin-independent (SI) and spin-dependent (SD) DM-proton scattering, as well as DM-electron scattering, comparing the predicted annihilation fluxes against neutrino~\cite{Kappl:2011kz, Bernal:2012qh, IceCube:2025fcu, Hooper:2025ohk, Berlin:2024lwe, ANTARES:2016obx, Widmark:2017yvd, Catena:2016ckl, IceCube:2016yoy, Ng:1986qt, Krishna:2025ncv, Maity:2023rez, IceCube:2021xzo, Bell:2021esh, Bell:2011sn, Bell:2012dk, Kopp:2009et, Super-Kamiokande:2015xms, Nguyen:2025ygc, Nguyen:2026apa} and \mbox{$\gamma$-ray}~\cite{Leane:2017vag, HAWC:2018szf, HAWC:2022khj, Nisa:2019mpb, Bose:2021cou, Bell:2021pyy, Serini:2022aed, Andrade:2024ekx} observations. For several annihilation channels, the resulting constraints on the DM scattering cross section exceed the leading limits from terrestrial direct detection.

However, the SD DM-neutron interaction has never been directly constrained by solar observations. Earlier effective-operator analyses computed solar capture for the full set of nonrelativistic operators, in some cases including $^{3}$He among their targets~\cite{Catena:2015uha, Liang:2013dsa, Catena:2015iea, Vincent:2015gqa}, but derived limits only for isoscalar and isovector benchmark couplings, which are dominated by DM-proton scattering off hydrogen~\cite{Catena:2015iea, Feng:2011vu}, while dedicated SD studies restricted their solar capture calculations to hydrogen altogether~\cite{Liang:2013dsa}. Moreover, these earlier analyses considered only DM masses above $\sim$10~GeV and neglected evaporation. As a result, the neutron-only benchmark reported by direct detection experiments has never been extracted from solar observations. In this work, we compute both the capture and evaporation rates for this interaction for the first time, extending the solar constraints down to DM masses of 100~MeV.

The strongest current limits on the cross section come from xenon-based experiments~\cite{XENON:2025vwd, LZ:2024zvo, PandaX:2024qfu}, whose sensitivity relies on the large natural abundances of $^{129}$Xe and $^{131}$Xe, which have an odd number of neutrons~\cite{Giffin:2025hdx, Menendez:2012tm, Freytsis:2010ne, Agrawal:2010fh, Tovey:2000mm, Ema:2024oce}. The target property that makes xenon so sensitive, however, works against the Sun: SD scattering proceeds only off nuclei with a non-zero neutron spin expectation value~\cite{Engel:1989ix, Ellis:1987sh, Pacheco:1989jz, Engel:1992bf, Divari:2000dc, Bednyakov:2004xq}, and for solar targets, this restricts the target population to odd-neutron isotopes such as $^{3}$He, $^{13}$C, $^{17}$O, $^{21}$Ne, $^{25}$Mg, and $^{29}$Si. These isotopes are present only in trace abundances~\cite{Asplund:2009fu}, orders of magnitude below the hydrogen and helium-4 that dominate SI and SD DM-proton capture, which strongly suppresses the SD DM-neutron capture rate and has left this channel unexplored. Nevertheless, this case is not hopeless. The dominant odd-neutron target $^{3}$He still reaches as much as 0.1\% of the hydrogen density at its peak, and the solar neutrino and $\gamma$-ray observations described above have improved dramatically over the past decade, providing a first opportunity to explore this interaction.

Motivated by this gap in the solar capture scenario, in this paper we calculate in detail the capture and evaporation of DM particles scattering with odd-neutron isotopes in the standard solar model, AGSS09~\cite{Asplund:2009fu}, through the SD DM-neutron interaction. We find that the accumulated DM population, though smaller than in the previous proton and electron scattering cases, still yields sufficient annihilation fluxes to be observed by current solar neutrino and $\gamma$-ray observations. Using current neutrino observations from Super-K and IceCube, as well as $\gamma$-ray observations from Fermi-LAT, HAWC, and LHAASO, we derive for the first time the astrophysical constraints and projections on the SD DM-neutron cross section. We also project the reach of future experiments such as Hyper-K, IceCube Upgrade, and SWGO. We find that, for some DM annihilation channels, these constraints and projections surpass current direct detection limits and can even reach into the neutrino-fog regime.

This paper is organized as follows. In Sec.~\ref{sect:SDscattering}, we discuss the solar capture and evaporation rate calculations for the SD DM-neutron interaction. In Sec.~\ref{sect:Annihilate}, we compute the captured DM annihilation rate, as well as the neutrino and $\gamma$-ray spectra for different annihilation channels. In Sec.~\ref{sect:obser}, we discuss all current and future neutrino and $\gamma$-ray observations that can be sensitive to these DM annihilation signals, as well as the strategies to probe the cross section for each observation. In Sec.~\ref{sect:constraints}, we derive the constraints for several optimistic annihilation channels and demonstrate the advantage of solar observations in probing the SD DM-neutron cross section. Finally, we conclude in Sec.~\ref{sect:conclusion}.

\section{Solar Spin-Dependent Dark Matter-Neutron Scattering}
\label{sect:SDscattering}

DM particles can scatter off any solar nucleus $A_{i}$ that carries a total angular momentum $J_{i}$. The corresponding spin-dependent DM-nucleus cross section is
\begin{equation}
    \sigma^{\rm SD}_{\chi A_{i}}=\left( \frac{\tilde{\mu}_{A_{i}}}{\tilde{\mu}_{n}} \right)^{2}\frac{4(J_{i}+1)}{3J_{i}}\,|\langle S_{n,i}\rangle|^{2}\,\sigma_{\chi n}^{\rm SD},
    \label{eq:sigmaSD}
\end{equation}
where $\tilde{\mu}_{A_{i}}$ and $\tilde{\mu}_{n}$ are the reduced masses of the DM-nucleus $i$ and the DM-neutron systems, $\langle S_{n,i}\rangle$ is the expectation value of the neutron spin in nucleus $i$, and $\sigma_{\chi n}^{\rm SD}$ is the spin-dependent DM-neutron cross section. Note that for nuclei with J$_i$ = 0, we also have $\langle S_{n,i}\rangle$ = 0 and the spin-dependent cross section vanishes. We adopt expectation values for the neutron spin from Refs.~\cite{Engel:1989ix, Ellis:1987sh, Pacheco:1989jz, Engel:1992bf, Divari:2000dc, Bednyakov:2004xq}. These expectation values carry nuclear-structure uncertainties, which are mild for $^{3}$He but can reach a factor of $\sim$2 in $\langle S_{n}\rangle^{2}$ for the heavier isotopes.

Equation~\eqref{eq:sigmaSD} follows the standard direct-detection convention of reporting SD limits in terms of effective single-nucleon cross sections, usually under proton-only or neutron-only coupling assumptions~\cite{Tovey:2000mm,PandaX:2025rrz}. Phenomenologically, this benchmark can be generated by axial interactions of the form 
\begin{equation}
    \mathcal{L}\supset\sum_{N=p,n} c_N(\bar{\chi}\gamma_\mu\gamma^5\chi)(\bar{N}\gamma^\mu\gamma^5N),
\end{equation}
whose leading nonrelativistic limit contains the standard spin-dependent operator \mbox{$\mathcal{O}_4=\vec S_\chi\cdot\vec S_N$}~\cite{Fitzpatrick:2012ix}. Following the usual direct-detection notation, the proton and neutron coefficients of \mbox{$\mathcal{O}_4^{(p)}$} and \mbox{$\mathcal{O}_4^{(n)}$} can be rewritten in terms of isospin-respecting and isospin-violating combinations, $a_0=a_n+a_p$ and $a_1=a_p-a_n$, respectively~\cite{Fitzpatrick:2012ix}. The neutron-only limit corresponds to $a_p=0$, or equivalently $a_1=-a_0$, up to the overall normalization convention. 

This scenario is particularly relevant for xenon direct detection because the naturally abundant spinful isotopes $^{129}$Xe and $^{131}$Xe have unpaired neutrons, giving xenon much stronger sensitivity to SD-neutron than SD-proton interactions~\cite{XENON100:2013ele}. We note that there are other operators that can generate spin-dependent interactions when the DM is a scalar, or fermionic with a vector DM-mediator coupling~\cite{Liang:2013dsa, Catena:2015uha, Vincent:2015gqa}. However, in these cases, the scattering cross section will have non-trivial momentum dependence, which is beyond the scope of this calculation~\cite{Berlin:2014tja}.

The relation in Eq.~\eqref{eq:sigmaSD} shows that only nuclei with non-zero spin, $J_{i}\neq 0$, and a non-vanishing neutron spin expectation value contribute to SD DM-neutron scattering, in contrast to the spin-independent and SD DM-proton cases dominated by hydrogen~\cite{Boddy:2022tyt}. In the following, we first present the number density profiles of these odd-neutron isotopes in the Sun (Sec.~\ref{ssect:profile}), and then compute the resulting solar capture and evaporation rates for SD DM-neutron scattering (Sec.~\ref{ssect:capt_eva}).

\subsection{Odd-Neutron Isotope Density Profiles in the Sun}
\label{ssect:profile}

\begin{figure}[tb]
    \centering
    \includegraphics[width=1\linewidth]{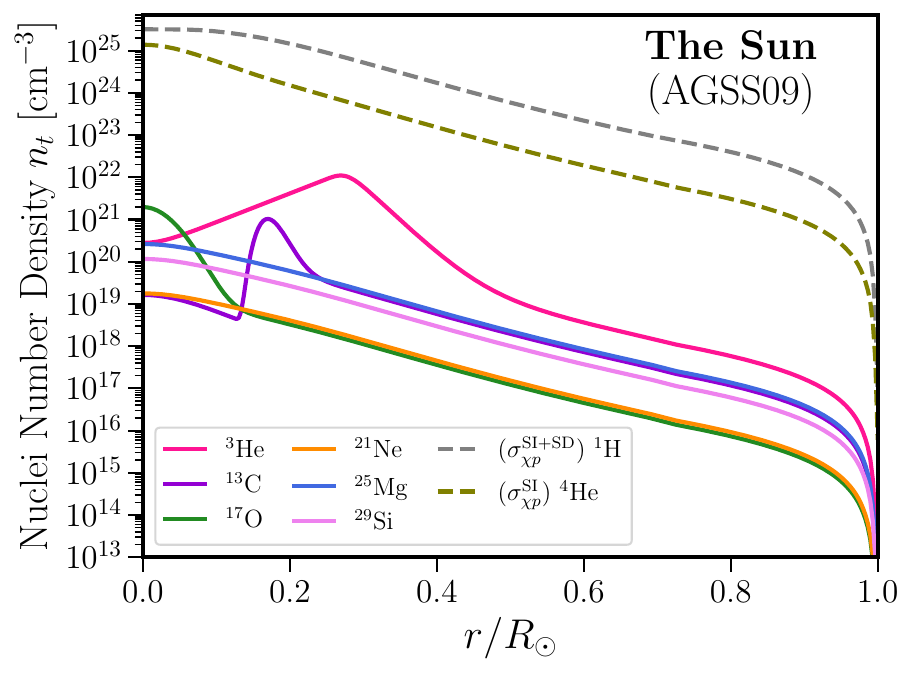}
    \caption{Number densities of the nuclei with non-negligible average neutron spin in the Sun, using the AGSS09 solar model~\cite{Asplund:2009fu}. The odd-neutron isotope densities relevant for SD DM-neutron scattering are shown as solid lines: $^{3}$He (magenta), $^{13}$C (purple), $^{17}$O (green), $^{21}$Ne (orange), $^{25}$Mg (blue), and $^{29}$Si (violet). For comparison, the dominant targets for DM-proton scattering are shown as dashed lines: $^{1}$H (gray), which contributes to both the SI and SD interactions, and $^{4}$He (olive), which contributes to the SI interaction only.}
    \label{fig:profiles}
\end{figure}

To identify the odd-neutron isotopes available for SD DM-neutron scattering, we use the AGSS09 standard solar model from Ref.~\cite{Asplund:2009fu} and retain all odd-neutron isotopes with a non-zero neutron spin expectation value~\cite{Bednyakov:2004xq}. These fall into two classes. The first comprises intermediates of hydrogen burning, whose abundances are governed by nuclear-reaction equilibrium rather than by a fixed isotopic fraction:

\begin{itemize}
    \item \textbf{$^{3}$He} ($J=1/2$): an intermediate of the $pp$-chain, produced by deuterium burning and destroyed in subsequent $^{3}$He-burning reactions. Because $^{3}$He burning is strongly temperature-sensitive, it is consumed in the hot core but survives at larger radii, so that its abundance peaks at intermediate radius ($r\simeq0.27\,R_{\odot}$), reaching $\sim$0.1\% of the hydrogen density.
    \item \textbf{$^{13}$C and $^{17}$O} ($J=1/2$ and $5/2$): intermediates of the carbon-nitrogen-oxygen CNO cycle. Their abundances peak or fall sharply at the inner radii where the corresponding proton-capture reactions reach equilibrium, producing the non-monotonic profiles seen in Fig.~\ref{fig:profiles}.
\end{itemize}
The second class comprises trace odd-neutron isotopes of heavier elements, present at fixed isotopic fractions and declining monotonically from the core outward:
\begin{itemize}
    \item \textbf{$^{21}$Ne} ($J=3/2$): the odd-neutron isotope of neon, with a solar isotopic abundance of $\sim 0.27\%$.
    \item \textbf{$^{25}$Mg} ($J=5/2$): the only stable odd-mass isotope of magnesium, with a solar isotopic abundance of $\sim 10\%$.
    \item \textbf{$^{29}$Si} ($J=1/2$): the odd-neutron isotope of silicon, with a solar isotopic abundance of $\sim 4.7\%$.
\end{itemize}

\begin{table}[tb]
\caption{Ranges give the spread in the $\langle S_{n}\rangle$ values across the nuclear-model variants (OGM, EOGM, shell model) compiled in Ref.~\cite{Bednyakov:2004xq}. Properties of the odd-neutron solar targets: nuclear spin $J$, neutron spin expectation value $\langle S_{n}\rangle$~\cite{Engel:1989ix, Ellis:1987sh, Pacheco:1989jz, Engel:1992bf, Divari:2000dc, Bednyakov:2004xq}, and isotopic fraction of the parent element. The abundances of the hydrogen-burning intermediates are set by reaction equilibrium rather than by a fixed fraction.}
\begin{ruledtabular}
\begin{tabular}{lccc}
Isotope & $J$ & $\langle S_{n}\rangle$ & Isotopic fraction \\
\hline
$^{3}$He & $1/2$ & $0.46$--$0.56$ & $pp$-chain equilibrium \\
$^{13}$C & $1/2$ & $-0.18$ & CN-cycle equilibrium \\
$^{17}$O & $5/2$ & $0.50$ & CNO-cycle equilibrium \\
$^{21}$Ne & $3/2$ & $0.17$--$0.29$ & $0.27\%$ \\
$^{25}$Mg & $5/2$ & $0.22$--$0.38$ & $10\%$ \\
$^{29}$Si & $1/2$ & $0.13$--$0.20$ & $4.7\%$ \\
\end{tabular}
\end{ruledtabular}
\label{tab:targets}
\end{table}

We do not include $^{57}$Fe, the odd-neutron isotope of iron, among our targets. Its ground state has an anomalously small magnetic moment, indicating that the nuclear spin is not carried predominantly by the valence neutrons~\cite{HAMAMOTO1962457}, so that its neutron spin expectation value is suppressed and is not reliably tabulated in the spin-structure literature. Combined with its low solar abundance ($\sim 2.1\%$ of iron), $^{57}$Fe contributes negligibly to the SD DM-neutron capture rate, and we conservatively omit it.

Figure~\ref{fig:profiles} shows the number densities of the odd-neutron isotopes we consider, using the AGSS09 solar model. Among these targets, $^{3}$He dominates throughout the solar interior,  exceeding the densities of the other odd-neutron isotopes by one to two orders of magnitude and reaching a peak of $\sim 10^{22}~{\rm cm}^{-3}$ at $r \simeq 0.27\,R_{\odot}$, where the temperature-dependent equilibrium between its $pp$-chain production and destriction is largest. The two CNO intermediates are more sharply structured. $^{13}$C is enhanced by more than an order of magnitude in a narrow shell near $r \simeq 0.17\,R_{\odot}$, reaching $\sim 10^{21}~{\rm cm}^{-3}$, where $^{12}$C$(p,\gamma)$ has already driven the $^{12}$C/$^{13}$C ratio to its equilibrium value while the much slower $^{14}$N$(p,\gamma)$ has yet to convert the carbon into $^{14}$N; interior to this shell the full CN cycle reaches equilibrium and $^{13}$C is instead depleted by an order of magnitude. $^{17}$O, whose parent $^{16}$O is never appreciably consumed, is monotonic: it peaks at the center at $\sim 2 \times 10^{21}~{\rm cm}^{-3}$, roughly two orders of magnitude above its primordial isotopic fraction, and falls steeply to that fraction beyond $r \simeq 0.15\,R_{\odot}$.  The remaining targets, $^{21}$Ne, $^{25}$Mg and $^{29}$Si, are fixed trace fractions of their parent elements and fall monotonically from the core outward. For comparison, we also show $^{1}$H and $^{4}$He, the dominant targets of SD and SI DM-proton scattering: even at its peak, the $^{3}$He density lies three to four orders of magnitude below that of $^{1}$H, setting the scale of the capture and evaporation rate suppression for the SD DM-neutron interaction.

\subsection{Solar Capture and Evaporation Rates}
\label{ssect:capt_eva}

We follow the conventional calculation of DM scattering inside the solar medium from Refs.~\cite{Gould:1987ju, Garani:2017jcj, Busoni:2017mhe}. Assuming that the DM-neutron scattering is isotropic and momentum-independent, as is conventionally adopted for nucleon scattering in direct detection, the rate for a DM particle with initial velocity $w$ to scatter off a nucleus $A_{i}$ to a final velocity $v$ is
\begin{align}
\label{eq:Rpm}
    R^{\pm}_{i}(w\to v)&=\frac{2}{\sqrt{\pi}}\frac{\mu_{i,+}^{2}}{\mu_{i}}\frac{v}{w}n_{i}(r)\sigma_{\chi A_{i}}\\
    \times&\left[ \chi(\pm \alpha_{-},\alpha_{+}) + \chi(\pm \beta_{-},\beta_{+})e^{\mu_{i}(w^{2}-v^{2})/u_{i}^{2}(r)} \right],\nonumber
\end{align}
where $n_{i}(r)$ is the number density of target nuclei $A_{i}$, shown in Fig.~\ref{fig:profiles}, and $\sigma_{\chi A_{i}}$ is the DM-nucleus cross section of Eq.~\eqref{eq:sigmaSD}. The ``$-$'' sign is for the case $w>v$ (capture), while the ``$+$'' sign is for the case $w<v$ (evaporation). We follow the notation of Ref.~\cite{Garani:2017jcj} for these functions as
\begin{align}
    &\mu_{i} \equiv m_{\chi}/m_{i},\,\, \mu_{i, \pm} \equiv (\mu_{i}\pm 1)/2,\\
    &u_{i}(r) \equiv \sqrt{2 T_{\odot}(r)/m_{i}},\\
    &\chi (a,b)\equiv \int_{a}^{b} e^{-y^{2}}{\rm d}y,\\
    & \alpha_{\pm}\equiv (\mu_{i,+}v\pm \mu_{i,-}w)/u_{i}(r),\\
    & \beta_{\pm} \equiv (\mu_{i,-}v\pm \mu_{i,+}w)/u_{i}(r),
\end{align}
where $m_{\chi}$ and $m_{i}$ are the DM and nuclei masses, and $T_{\odot}(r)$ is the solar temperature profile.

\begin{figure*}[tb]
    \centering
    \includegraphics[width=1\columnwidth]{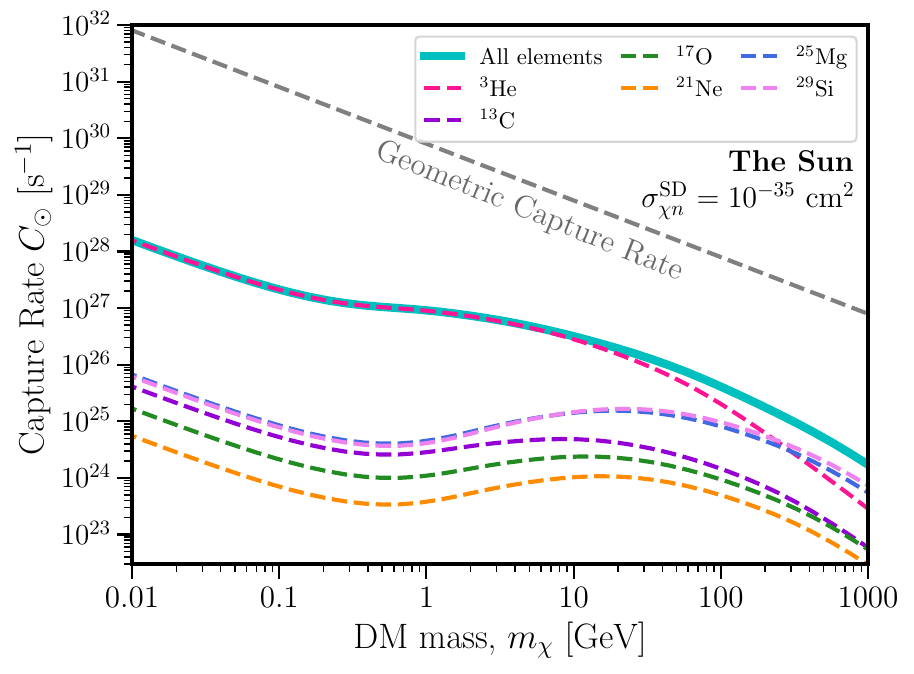}
    \hfill
    \includegraphics[width=1\columnwidth]{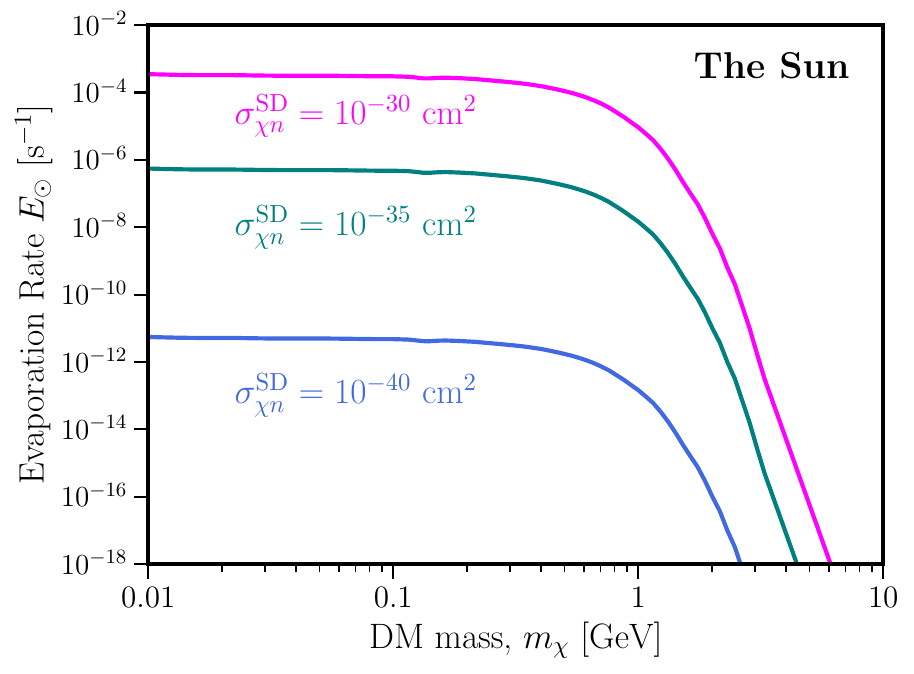}
    \caption{{\bf Left:} Capture rate for a spin-dependent DM-neutron cross section of $\sigma^{\rm SD}_{\chi n}=10^{-35}$~cm$^{2}$. The total capture rate, including all odd-neutron isotopes, is shown in teal, and contributions from each elements are in dashed lines: $^{3}$He (magenta), $^{13}$C (purple), $^{17}$O (green), $^{21}$Ne (orange), $^{25}$Mg (blue), and $^{29}$Si (violet). The solar geometric (maximum) capture rate is the dashed gray line. {\bf Right:} The evaporation rate for the same interaction, for three benchmark cross sections: $10^{-40}$~cm$^{2}$ (blue), $10^{-35}$~cm$^{2}$ (teal), and $10^{-30}$~cm$^{2}$ (magenta).}
    \label{fig:capt_eva}
\end{figure*}

In the DM capture scenario, halo DM particles travel toward the Sun and fall into its gravitational potential. These DM particles can then scatter off the odd-neutron isotopes inside the Sun and lose kinetic energy. If the DM velocity falls below the Sun's escape velocity, the particle is captured. In the optically thin limit, where DM interacts with the SM very weakly, the capture rate is
\begin{equation}
    \begin{split}
    C_{\rm weak}=&\int_{0}^{R_{\odot}}{\rm d}r\, 4\pi r^{2}\int_{0}^{\infty}{\rm d}u_{\chi}\left( \frac{\rho_{\chi}}{m_{\chi}}\right) \frac{f_{v_{\odot}}(u_{\chi})}{u_{\chi}}\\
    &\times \int_{0}^{v_{e}(r)}{\rm d}v\, w(r)\sum_{i}R_{i}^{-}(w\to v)|F_{i}(q)|^{2},
    \label{eq:Cweak}
    \end{split}
\end{equation}
where we sum over all target isotopes for SD DM-neutron scattering. We adopt a solar radius of \mbox{$R_{\odot}=696340$~km} and a local solar DM density of \mbox{$\rho_{\chi}=0.4$~GeV/cm$^{3}$}. The DM speed in the solar gravitational potential prior to scattering depends on its initial velocity $u_{\chi}$ and the solar escape velocity $v_{e}(r)$ as
\begin{equation}
w(r)=\sqrt{u_{\chi}^{2}+v_{e}^{2}(r)}.
\end{equation}
The halo DM velocity distribution is taken from the Standard Halo Model as
\begin{equation}
    f_{v_{\odot}}(u_{\chi})=\sqrt{\frac{3}{2\pi}}\frac{u_{\chi}}{v_{\odot}v_{d}}\left( e^{-\frac{3(u_{\chi}-v_{\odot})^{2}}{2v_{d}^{2}}} - e^{-\frac{3(u_{\chi}+v_{\odot})^{2}}{2v_{d}^{2}}}\right),
\end{equation}
where the solar velocity \mbox{$v_{\odot}=220$~km/s} and the velocity dispersion \mbox{$v_{d}\simeq\sqrt{3/2}\,v_{\odot}=270$~km/s} are defined in the halo rest frame. Although the integral over $u_{\chi}$ formally extends to infinity, the velocity distribution falls off exponentially at high speeds and is kinematically bounded by the local Galactic escape velocity, \mbox{$v_{e}^{\rm gal}=533^{+54}_{-41}$~km/s} at 90\% confidence level~\cite{Piffl:2013mla}. We therefore truncate the integration at $u_{\chi}^{\rm max}=600$~km/s, although we note that raising the cutoff to $1000$~km/s leaves the capture rate unchanged.

For the SD DM-neutron interaction, we adopt a nuclear form factor corresponding to a Gaussian nuclear density distribution,
\begin{equation}
    |F_{i}(q)|^{2}=e^{-q^{2}r_{i}^{2}/3},
\end{equation}
with a root-mean-square radius $r_{i}$ that depends on the odd-neutron isotope as
\begin{equation}
\begin{split}
    r_{i}=\frac{\sqrt{3}}{2}\Big[ 1.7 A_{i}^{1/3}&-0.28 -0.78\Big( A_{i}^{1/3}-3.8\\
    &+\sqrt{(A_{i}^{1/3}-3.8)^{2}+0.2} \Big) \Big]~{\rm fm},
\end{split}
\end{equation}
and the transfer momentum for each scattering is approximated as:
\begin{equation}
    q^{2}=m_{i}m_{\chi}(w^{2}-v^{2}).
\end{equation}
Following the literature, we adopt a Gaussian form factor with an effective interaction radius for the nuclear spin distribution, in contrast to the density-based form factor, which is appropriate only for SI scattering~\cite{Engel:1991wq, Belanger:2008sj}.

The weak-interaction capture rate of Eq.~\eqref{eq:Cweak} grows linearly with the cross section, since it follows from Eqs.~\eqref{eq:sigmaSD} and \eqref{eq:Rpm}, in which $\sigma_{\chi A_{i}} \propto \sigma_{\chi n}^{\rm SD}$. This linear scaling cannot hold indefinitely: once the Sun becomes optically thick to DM scattering, every DM particle crossing the Sun is captured, setting an upper limit known as the geometric capture rate~\cite{Bernal:2012qh, Bottino:2002pd},
\begin{equation}
    C_{\rm geo}=\pi R_{\odot}^{2}\left( \frac{\rho_{\chi}}{m_{\chi}} \right)\langle v\rangle_{0}\left( 1 + \frac{3 v_{e}^{2}(R_{\odot})}{2 v_{d}^{2}} \right)\xi_{\odot},
\end{equation}
where $\langle v\rangle_{0}=\sqrt{8/(3\pi)}\,v_{d}$ is the mean DM speed in the halo rest frame, the factor $\left(1+3v_{e}^{2}(R_{\odot})/2v_{d}^{2}\right)$ accounts for gravitational focusing by the Sun, and $\xi_{\odot}\simeq 0.81$ accounts for the suppression due to the solar motion through the halo. We then interpolate between the single-scatter capture rate of Eq.~\eqref{eq:Cweak} and this geometric limit following Ref.~\cite{Bernal:2012qh} as
\begin{equation}
    C_{\odot}=C_{\rm weak}\left( 1 - e^{-C_{\rm geo}/C_{\rm weak}} \right).
\end{equation}
Figure~\ref{fig:capt_eva} (left panel) shows the resulting capture rates for the SD DM-neutron interaction at a benchmark cross section of \mbox{$\sigma^{\rm SD}_{\chi n}=10^{-35}$~cm$^{2}$}. We compute the rate in two cases: first for each target element individually, and then combining all six odd-neutron isostopes. We find that the total capture rate and the $^{3}$He-only capture rate are nearly identical below $\sim 40$~GeV, where $^{3}$He dominates the capture rate, both because it is the most abundant of our targets and because its small mass provides the best kinematic target for light DM. At higher masses, the heavier isotopes contribute an increasing fraction of the total rate, and the full calculation rises above the $^{3}$He-only result.

A further feature of the SD DM-neutron interaction is that the capture rate is suppressed relative to the traditional SI and SD DM-proton and DM-electron cases by roughly 4--5 orders of magnitude~\cite{Garani:2017jcj, Busoni:2017mhe, Nguyen:2025ygc}. This suppression directly reflects the scarcity of the targets: even the dominant $^{3}$He is 3--5 orders of magnitude less abundant than $^{1}$H and $^{4}$He (Fig.~\ref{fig:profiles}). As a consequence, the saturation (transition) cross section, above which the Sun becomes optically thick and the capture rate reaches its geometric limit~\cite{Leane:2023woh}, is pushed from the canonical value of $\sim 10^{-35}$~cm$^{2}$ up to $\sim 10^{-31}$~cm$^{2}$, a distinctive feature of the SD DM-neutron interaction in the Sun.

\begin{figure*}[tb]
    \centering
    \includegraphics[width=1\linewidth]{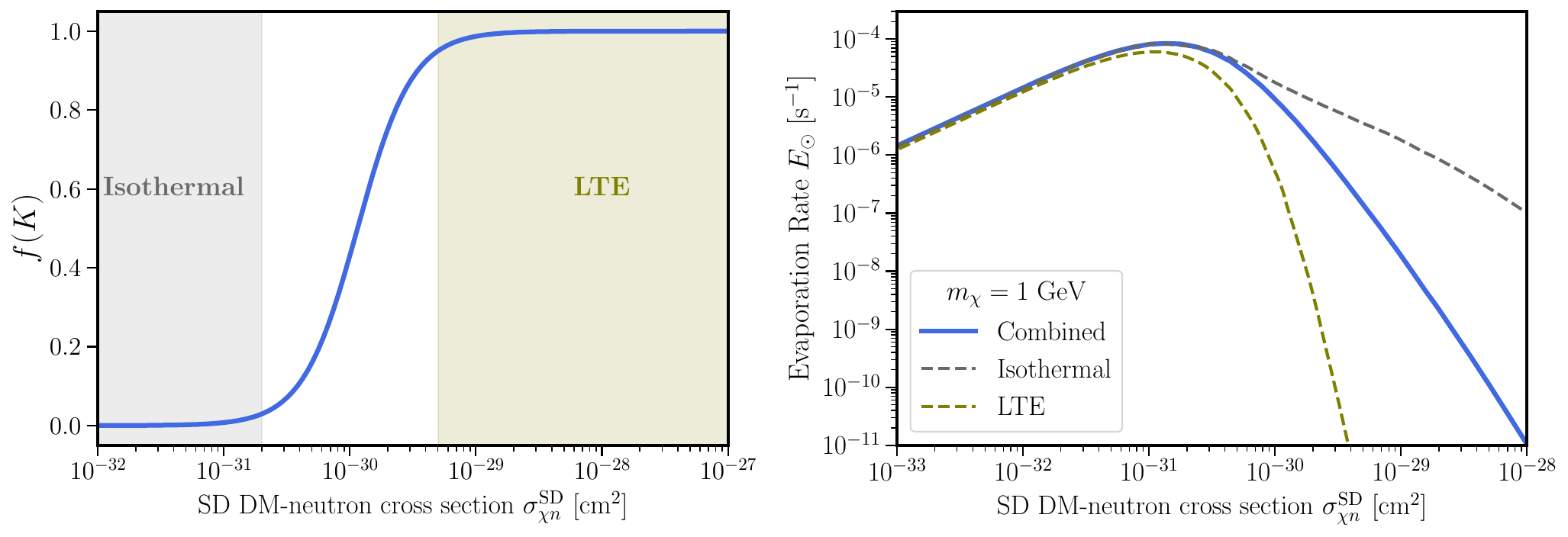}
    \caption{{\bf Left:} $f(K)$ as a function of the SD DM-neutron cross section. The gray and olive shades are for the isothermal and LTE dominated regimes. {\bf Right:} The evaporation rate for three different treatments of the captured DM number density, for a benchmark DM mass of $m_{\chi}=1$~GeV: The dashed gray line is for the isothermal profile, and the dashed olive line is for the LTE profile. The solid blue line shows the result for the combined profile, using the fraction $f(K)$ as a function of the SD DM-neutron cross section.}
    \label{fig:iso_lte}
\end{figure*}

In the opposite process, captured DM particles can be upscattered by the odd-neutron nuclei inside the Sun and gain kinetic energy. If their final velocity exceeds the local escape velocity, they evaporate. The evaporation rate is given by
\begin{align}
    E_{\odot}&=\sum_{i}\int_{0}^{R_{\odot}}{\rm d}r\, 4\pi r^{2}s(r)n_{\chi}(r)\label{eq:EvaRate}\\
    &\times\int_{0}^{v_{e}(r)}{\rm d}w\, 4\pi w^{2}f_{\chi}(\boldsymbol{w}, r)\int_{v_{e}(r)}^{\infty}{\rm d}v\,R_{i}^{+}(w\to v),\nonumber
\end{align}
where we model the captured DM velocity distribution as a Maxwell--Boltzmann distribution truncated at the local escape velocity,
\begin{equation}
    f_{\chi}(\boldsymbol{w}, r)=\frac{e^{-w^{2}/v_{\chi}^{2}(r)}\,\Theta(v_{e}(r)-w)}{\sqrt{\pi^{3}}\,v_{\chi}^{3}(r)\left[ {\rm Erf}\left(\frac{v_{e}}{v_{\chi}}\right) - \frac{2v_{e}}{\sqrt{\pi}\,v_{\chi}}e^{-v_{e}^{2}/v_{\chi}^{2}}\right]},
\end{equation}
with a thermal speed set by the DM temperature,
\begin{equation}
    v_{\chi}(r)=\sqrt{2T_{\chi}(r)/m_{\chi}}.
\end{equation}
The DM temperature $T_{\chi}(r)$ depends on the SD DM-neutron cross section, which determines the thermalization regime of the captured population~\cite{Gould:1989hm, Banks:2021sba}:
\begin{itemize}
    \item \textbf{Isothermal regime (weak cross section, optically thin).} When the cross section is small, the mean free path of a captured DM particle is large compared to the size of the captured distribution, so each particle samples the entire solar interior between scatters and reaches a single, position-independent temperature. The DM distribution is then characterized by one global temperature, $T_{\chi}(r)=T_{\chi}^{\rm iso}$, which we obtain following Refs.~\cite{Blanco:2024lqw, Blanco:2025wpo}
    \begin{equation}
        T_{\chi}^{\rm iso}=\frac{T_{\odot}(0)+T_{\odot}(\min(r_{\chi}, R_{\odot}))}{2},
    \end{equation}
    where the DM scale radius $r_{\chi}$ is defined as
    \begin{equation}
        r_{\chi}=\sqrt{\frac{3T_{\odot}(0)}{2\pi G_{N}\rho_{\odot}(0)m_{\chi}}},
    \end{equation}
    where $\rho_{\odot}$ is the solar mass density and $G_{N}$ is Newton's constant.
    \item \textbf{Local thermal equilibrium (LTE) regime (strong cross section, optically thick).} When the cross section is large, the mean free path is short and a captured DM particle scatters many times within a small region, thermalizing locally with the surrounding nuclei. The DM temperature then tracks the solar temperature at each radius, $T_{\chi}(r)=T_{\odot}(r)$.
\end{itemize}

These two regimes also determine how the captured DM is distributed inside the Sun whenever the DM particles thermalize with the solar medium. We find that this thermalization condition is satisfied for the range of SD DM-neutron interactions that are relevant for the constraints set in this paper. In the isothermal regime, the DM distribution is
\begin{equation}
    n_{\chi}^{\rm iso}(r) = N_{\rm iso}\, e^{-m_{\chi}\phi(r)/T_{\chi}},
\end{equation}
with the gravitational potential
\begin{equation}
    \phi(r)=\int_{0}^{r}{\rm d}r^{\prime}\, G_{N}M_{\odot}(r^{\prime})/r^{\prime 2}.
\end{equation}
In the LTE regime, the DM number density is instead given by~\cite{Gould:1989hm, Nauenberg:1986em, Banks:2021sba}
\begin{align}
        n_{\chi}^{\rm LTE}(r)&=N_{\rm LTE}\left[ \frac{T_{\odot}(r)}{T_{\odot}(0)} \right]^{3/2}\\
        &\times \exp\left[ - \int _{0}^{r}{\rm d}r^{\prime}\frac{\alpha(r^{\prime})\frac{{\rm d}T_{\odot}(r^{\prime})}{{\rm d}r^{\prime}}+m_{\chi}\frac{{\rm d}\phi(r^{\prime})}{{\rm d}r^{\prime}}}{T_{\odot}(r^{\prime})} \right],\nonumber
\end{align}
where $\alpha(r)$ is the dimensionless thermal diffusivity, which accounts for all scattering targets as
\begin{equation}
    \alpha(r) = \ell(r)\sum_{i}\ell_{i}^{-1}(r)\,\alpha_{0}(\mu_{i}),
\end{equation}
and we adopt the single-target thermal diffusivity $\alpha_{0}(\mu_{i})$ from Ref.~\cite{Leane:2022hkk}. The total and partial mean free paths are defined as
\begin{equation}
    \ell^{-1}(r)=\sum_{i}\ell_{i}^{-1}(r),\qquad \ell_{i}(r)=\frac{1}{\sigma_{\chi A_{i}}n_{i}(r)}.
\end{equation}
The normalization constants $N_{\rm iso}$ and $N_{\rm LTE}$ are fixed by requiring
\begin{equation}
    \int_{0}^{R_{\odot}}{\rm d}r \, 4\pi r^{2}\, n_{\chi}^{\rm iso/LTE}(r)=1.
\end{equation}

For an arbitrary SD DM-neutron cross section, the DM number density is an interpolation between the two regimes~\cite{Bottino:2002pd, Scott:2008ns, Garani:2017jcj},
\begin{equation}
    n_{\chi}(r)=f(K)\,n_{\chi}^{\rm LTE}(r)+[1 - f(K)]\,n_{\chi}^{\rm iso}(r),
    \label{eq:nXtotal}
\end{equation}
where $f(K)$ is a function of the Knudsen number $K$,
\begin{equation}
    K\equiv \frac{\ell(0)}{r_{\chi}},\qquad f(K)=\frac{1}{1 + (K/K_{0})^{2}},
\end{equation}
and $K_{0}=0.4$ is the value at which energy transport by DM particles is most efficient~\cite{Gould:1989hm}. The Knudsen number compares the DM mean free path to the scale radius $r_{\chi}$: in the isothermal regime the mean free path is long and $K\gg 1$, while in the LTE regime it is short and $K\ll 1$.

Finally, the suppression factor $s(r)$ is given by
\begin{equation}
    s(r) = \eta_{\rm ang}(r)\, \eta_{\rm mult}(r)\, e^{-\tau(r)},
    \label{eq:s(r)}
\end{equation}
where $\tau(r)$ is the optical depth in the radial direction
\begin{equation}
    \tau(r) = \int_{r}^{R_{\odot}}{\rm d}r^{\prime}\,\ell^{-1}(r^{\prime}).
\end{equation}
We follow Ref.~\cite{Garani:2017jcj} for the remaining quantities,
\begin{align}
    \eta_{\rm ang}(r)&=\frac{7}{10}\frac{1 - e^{-10\tau(r)/7}}{\tau(r)},\\
    \eta_{\rm mult}(r)&={}_{0}F_{1}\big(;\,1 + \frac{2}{3}\hat{\phi}(r);\,\tau(r)\big),
\end{align}
where $_{0}F_{1}(;b;z)$ is the confluent hypergeometric limit function, and
\begin{equation}
    \hat{\phi}(r)\equiv \frac{m_{\chi}v_{e}^{2}(r)}{2T_{\odot}(r)}
\end{equation}
is the local dimensionless escape energy.

Figure~\ref{fig:capt_eva} (right panel) shows the evaporation rate for three benchmark SD DM-neutron cross sections. As with the capture rate, the evaporation rate is suppressed relative to the SD DM-proton case by roughly 4 orders of magnitude, reflecting the low abundance of odd-neutron isotopes compared to hydrogen in the Sun. Nevertheless, the rate still falls below the inverse age of the Sun ($\sim 10^{-18}~{\rm s}^{-1}$) only around $2$--$4$~GeV for cross sections at or below $10^{-35}$~cm$^{2}$, indicating that the evaporation mass in the SD DM-neutron case is similar to that of the proton-scattering case.

Notably, the evaporation rate does not scale linearly with the cross section. As shown in Fig.~\ref{fig:iso_lte} (left panel), the transition from the isothermal to the LTE regime occurs for cross sections larger than $\sim 10^{-31}$~cm$^{2}$. At these high cross sections, the captured DM distribution becomes centrally concentrated and the Sun turns optically thick, so that a DM particle upscattered above the escape velocity is likely to rescatter and re-cool before leaving the Sun. This effect is captured by the suppression factor $s(r)$ in Eqs.~\eqref{eq:EvaRate} and \eqref{eq:s(r)}, which reduces the fraction of upscattered particles that successfully evaporate and becomes important precisely in this regime. Figure~\ref{fig:iso_lte} (right panel) shows the resulting evaporation rate for a benchmark mass $m_{\chi}=1$~GeV, computed using the isothermal profile, the LTE profile, and the interpolated distribution of Eq.~\eqref{eq:nXtotal}. Beyond the isothermal--LTE transition, the suppression drives the evaporation rate down with increasing cross section, in contrast to its linear growth at low cross sections. Because a lower evaporation rate allows lighter DM to survive against evaporation, this suppression shifts the evaporation mass to lower values at large cross sections~\cite{Garani:2017jcj, Busoni:2017mhe, Leane:2022hkk, Nguyen:2026nhe}. 

\section{Dark Matter Annihilation and Signals}
\label{sect:Annihilate}

Captured DM particles can annihilate with one another into SM final states. The annihilation rate for a pair of DM particles is
\begin{equation}
    A_{\odot}=\langle \sigma v\rangle_{\chi\chi}\frac{\int_{0}^{R_{\odot}}{\rm d}r\, 4\pi r^{2}n_{\chi}^{2}(r)}{\left( \int_{0}^{R_{\odot}}{\rm d}r\, 4\pi r^{2} n_{\chi}(r)\right)^{2}},
\end{equation}
where we assume a thermally averaged $s$-wave cross section \mbox{$\langle \sigma v\rangle_{\chi\chi}=3\times 10^{-26}$~cm$^{3}$/s}. Captured DM particles reach equilibrium between capture and annihilation on a timescale
\begin{equation}
    t_{\rm eq}=\frac{1}{\sqrt{C_{\odot}A_{\odot}}}.
\end{equation}
To account for signals in regions of parameter space that lie outside of this equilibrium condition, we adopt the total annihilation rate of DM into SM particles, given by 
\begin{equation}
    \Gamma(m_{\chi}, \sigma^{\rm SD}_{\chi n})=\frac{C_{\odot}}{2}\left( \frac{\tanh(\kappa\, t_{\odot}/t_{\rm eq})}{\kappa + \frac{1}{2}E_{\odot}t_{\rm eq}\tanh(\kappa\, t_{\odot}/t_{\rm eq})} \right)^{2},
    \label{eq:Gamma}
\end{equation}
where \mbox{$\kappa = \sqrt{1 + (E_{\odot}t_{\rm eq}/2)^{2}}$} and the age of the Sun is \mbox{$t_{\odot}\simeq4.57$~Gyr}.

\begin{figure}[tb]
    \centering
    \includegraphics[width=1\columnwidth]{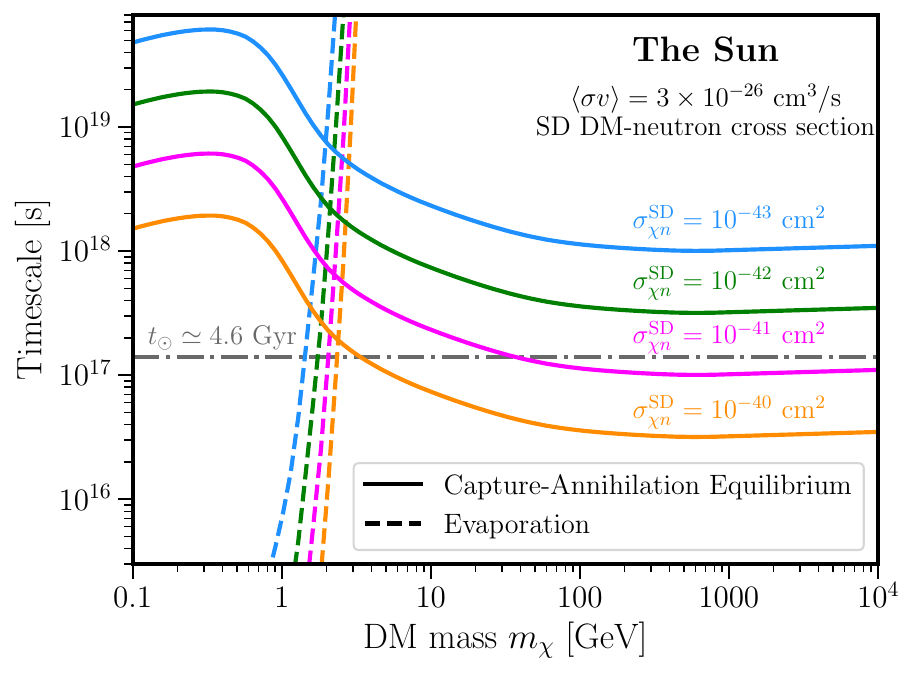}
    \caption{The solar equilibrium timescales between the DM capture and annihilation processes for several SD DM-neutron cross sections (solid lines), assuming a thermally averaged annihilation cross section $\langle\sigma v\rangle = 3\times10^{-26}$~cm$^{3}$/s. The evaporation timescales are shown as dashed lines, and the age of the Sun, $t_{\odot}\simeq 4.6$~Gyr, is indicated by the gray dash-dotted line.}
    \label{fig:timescale}
\end{figure}

Figure~\ref{fig:timescale} shows the capture-annihilation equilibrium timescale for the SD DM-neutron interaction at several benchmark cross sections (solid lines) and as a function of the DM mass, as well as the DM evaporation timescale, defined as the inverse of the evaporation rate $1/E_{\odot}$ (dashed lines), and the age of the Sun $t_{\odot}$ (dash-dotted line). The equilibrium between capture and annihilation is established within the solar lifetime only when the equilibrium timescale falls below $t_{\odot}$ and the evaporation timescale rises above it. We find that DM masses below $\sim 2$~GeV never reach equilibrium, as the evaporation rate becomes significant and rapidly depletes the captured population. The situation is also cross-section dependent: at $\sigma^{\rm SD}_{\chi n}\simeq 10^{-41}$~cm$^{2}$, only DM masses above $\sim 50$~GeV reach equilibrium, and for smaller cross sections equilibrium is never established within the solar lifetime. On the other hand, for $\sigma^{\rm SD}_{\chi n}$ above $\sim 10^{-40}$~cm$^{2}$ equilibrium holds across essentially the entire mass range. These results show that the equilibrium assumption cannot be applied blindly to the SD DM-neutron interaction, and any probe of this channel must carefully account for the interplay between capture, annihilation, and evaporation. In this work, we calculate the annihilation rate using Eq.~\eqref{eq:Gamma}, instead of the approximation $\Gamma=C/2$ from the equilibrium assumption, making our model accurate for arbitrary SD DM-neutron cross section values. 

The annihilation of these DM particles in the Sun can produce multiple signals accessible to current indirect-detection observations, such as neutrinos and $\gamma$-rays. The indirect-detection flux is given by
\begin{equation}
    \frac{{\rm d}\Phi}{{\rm d}E}\Big{|}_{\nu, \gamma}=\frac{\Gamma}{4\pi D_{\odot}^{2}}\frac{{\rm d}N}{{\rm d}E}\Big{|}_{\nu, \gamma}P_{\rm surv}^{\nu,\gamma},
\end{equation}
where $D_{\odot}=1$~AU is the distance from the Sun to detectors at Earth. The energy spectrum ${\rm d}N/{\rm d}E$ and the survival probability $P_{\rm surv}$ depend on the type of signal:
\begin{itemize}
    \item {\bf Neutrinos:} Owing to their weak coupling to the SM, neutrinos produced inside the Sun can easily escape. We consider two DM annihilation channels producing neutrinos: \mbox{$\chi\chi\to \nu\bar{\nu}$} and \mbox{$\chi\chi\to \tau^{+}\tau^{-}$}. The survival probability is~\cite{Nguyen:2025ygc, Nguyen:2026nhe, Nguyen:2026apa}
    \begin{equation}
        P_{\rm surv}^{\nu}(E_{\nu}) = \exp\left[ -\int_{0}^{R_{\odot}}{\rm d}r\, \sigma_{\nu {\rm H}}(E_{\nu})\, n_{\rm H}(r) \right],
    \end{equation}
    where $n_{\rm H}(r)$ is the number density of hydrogen, the dominant absorption target in the Sun, and $\sigma_{\nu {\rm H}}$ is the neutrino-hydrogen cross section~\cite{Zhou:2023mou}. For direct annihilation to a neutrino--anti-neutrino pair, the spectrum is
    \begin{equation}
        \frac{{\rm d}N_{\nu}}{{\rm d}E_{\nu}}\Big{|}_{\chi\chi\to\nu\bar{\nu}}=\frac{2}{3}\,\delta(E_{\nu}-m_{\chi}),
    \end{equation}
    where the factor $2/3$ accounts for neutrino oscillations and for the inability of the detectors to distinguish neutrinos from anti-neutrinos. In practice, this line spectrum is smeared by the detector energy resolution, which is normally modeled as Gaussian. For the $\tau^{+}\tau^{-}$ channel, we compute the spectrum using {\tt PPPC4DM}$\nu$~\cite{Baratella:2013fya} below 500~GeV and $\chi${\tt aro}$\nu$~\cite{Liu:2020ckq} for heavier DM masses. Notably, other channels such as $e^{+}e^{-}$ (through weak bremsstrahlung)~\cite{Maity:2023rez, Krishna:2025ncv}, $\mu^{+}\mu^{-}$, and $b\bar{b}$ can also produce neutrinos~\cite{Bernal:2012qh, Kappl:2011kz}. We restrict our analysis to the two channels above because they yield the brightest neutrino signals and are widely considered in both low- and high-energy neutrino observations.
    \item {\bf $\gamma$-rays:} DM can annihilate into long-lived particles $\phi$ that escape the Sun and subsequently decay into SM states, producing $\gamma$-ray signals~\cite{Leane:2017vag, Leane:2021ihh,Leane:2021tjj, Acevedo:2024ttq, Nguyen:2022zwb, Linden:2024uph}. The survival probability depends on the mediator lifetime $\tau_{\phi}$ as
    \begin{equation}
        P_{\rm surv}^{\gamma} = e^{-R_{\odot}/\eta c \tau_{\phi}} - e^{-D_{\odot}/\eta c\tau_{\phi}},
    \end{equation}
    where $\eta \simeq m_{\chi}/m_{\phi}$ is the boost factor, and $c$ is the speed of light. In our analysis, we adopt the most optimistic case $P_{\rm surv}^{\gamma}=1$. We consider the two most optimistic mediator decay channels. For \mbox{$\phi\to2\gamma$}, the spectrum is box-shaped~\cite{Abdullah:2014lla},
    \begin{equation}
        \frac{{\rm d}N_{\gamma}}{{\rm d}E_{\gamma}}\Big{|}_{\chi\chi\to4\gamma}=\frac{4}{\Delta E}\,\Theta(E-E_{\rm min})\,\Theta(E_{\rm max}-E),
    \end{equation}
    where \mbox{$\Delta E=\sqrt{m_{\chi}^{2}-m_{\phi}^{2}}$}, and the photon energies span
    \begin{equation}
        E_{\rm max/min}=\frac{m_{\chi}}{2}\left( 1 \pm \sqrt{1 - \frac{m_{\phi}^{2}}{m_{\chi}^{2}}}\right).
    \end{equation}
    We work in the limit $m_{\chi}\gg m_{\phi}$. We also consider the $\phi\to e^{+}e^{-}$ decay channel~\cite{Nguyen:2026nhe, Smolinsky:2017fvb, Feng:2016ijc}, in which photons are generated predominantly through final-state radiation and then boosted into the laboratory frame~\cite{Nguyen:2024kwy, Nguyen:2026nhe}. Notably, other SM decay channels of the long-lived mediator can also produce photon signals; however, previous studies have shown that they are suppressed compared to the two channels above~\cite{Leane:2017vag, Nguyen:2026nhe}. Since the purpose of this study is to identify the best prospects for current observations to probe the SD DM-neutron interaction, we do not consider these subdominant channels.
\end{itemize}

\section{Solar Neutrino and Gamma-Ray Observations}
\label{sect:obser}

We now discuss the potential observation of neutrino and $\gamma$-ray fluxes from DM annihilation in the Sun using current and upcoming solar observations. Neutrino signals are best probed by large-volume neutrino telescopes, with water Cherenkov detectors covering the low-energy regime and ice-based detectors extending the reach to higher DM masses. Potential $\gamma$-ray signals from long-lived mediator decays require detectors capable of observing the Sun at $\gamma$-ray energies, namely space-based telescopes and ground-based water Cherenkov or particle-array observatories. Notably, imaging atmospheric Cherenkov telescopes such as H.E.S.S.~\cite{HESS:2018pbp}, VERITAS~\cite{Weekes:2001pd}, MAGIC~\cite{Baixeras:2003xr}, and CTAO~\cite{CTAConsortium:2017dvg} cannot observe the Sun, since they operate only during astronomical darkness. We describe the neutrino observations in Sec.~\ref{ssect:neutrino_exp}, and the $\gamma$-ray observations in Sec.~\ref{ssect:gammaray_exp}, as well as the strategy we adopt to derive cross-section constraints from each.

\subsection{Neutrino Observations}
\label{ssect:neutrino_exp}
Solar neutrino observations have been a major focus of the neutrino physics community for decades. Measurements of the solar neutrino fluxes, from the dominant $pp$-chain components to the subdominant CNO cycle, have tested the standard solar model~\cite{Asplund:2009fu}, while the observed deficit of solar neutrinos led to the discovery of neutrino oscillations and the resolution of the solar neutrino problem~\cite{Davis:1964hf, Kamiokande-II:1989hkh, SNO:2002tuh}. Beyond these standard topics in neutrino astronomy, solar neutrino observations also provide a powerful probe of new physics, including the DM annihilation signals from the Sun considered in this work~\cite{Srednicki:1986vj}. Here, we consider four neutrino detectors: Super-K, Hyper-K, IceCube, and the IceCube~Upgrade.

\subsubsection{Super-Kamiokande and Hyper-Kamiokande}
\label{sssect:SKHK}

The Super-Kamiokande (Super-K) detector, situated roughly 1~km underground beneath Mount Ikeno in the Gifu Prefecture of Japan, has been a cornerstone of neutrino physics for nearly three decades~\cite{Super-Kamiokande:2002weg}. Super-K contains a fiducial mass of 22.5~kton of ultra-pure water and detects neutrinos through the Cherenkov light emitted by the charged particles produced in their interactions, delivering some of the most precise measurements of solar and atmospheric neutrinos available~\cite{Super-Kamiokande:2005wtt, Super-Kamiokande:2008ecj, Super-Kamiokande:2010tar}. Its successor, Hyper-Kamiokande (Hyper-K), currently under construction at a nearby site, will enlarge the fiducial mass to 187~kton, substantially extending the sensitivity to neutrinos in the MeV--GeV range~\cite{Abe:2011ts}.

For the neutrino fluxes from DM annihilation in the Sun and from the astrophysical background, we compute the expected number of observed events in the energy range $[E_{\nu}^{\rm min}, E_{\nu}^{\rm max}]$ as
\begin{equation}
    N_{\nu}^{\chi/{\rm bkg}}=\int_{E_{\nu}^{\rm min}}^{E_{\nu}^{\rm max}}\!{\rm d}E_{\nu}\,\frac{{\rm d}\Phi_{\nu}^{\chi/{\rm bkg}}}{{\rm d}E_{\nu}}\,\sigma_{\nu {\rm H}_{2}{\rm O}}(E_{\nu})\times \xi\,,
\end{equation}
\noindent where $\sigma_{\nu {\rm H}_2{\rm O}}$ is the neutrino-water cross section from Ref.~\cite{Zhou:2023mou} and $\xi=N_{{\rm H}_{2}{\rm O}}\times T_{\rm obs}$ is the detector exposure, with $N_{{\rm H}_{2}{\rm O}}$ the number of water molecules in the fiducial volume and $T_{\rm obs}$ the observation time. We consider only muon neutrinos, owing to their good angular reconstruction in water Cherenkov detectors, and adopt the fiducial atmospheric muon neutrino background consistent with the solar direction, as we previously calculated in Refs.~\cite{Nguyen:2025ygc, Nguyen:2026nhe, Nguyen:2026apa}. We calculate this fiducial background by cutting the all-sky atmospheric muon neutrino background in Ref.~\cite{Honda:2011nf}, noting that the detectors angular resolution varies from 90$^{\circ}$--2$^{\circ}$ for 0.1--100~GeV neutrino energies~\cite{Konishi:2010mv, Konishi:2011sc}. Throughout this work, we assume an exposure of $T_{\rm obs}=10$~years for both detectors, with fiducial masses of 22.5~kton for Super-K and 187~kton for Hyper-K. For the $\nu\bar{\nu}$ channel, we integrate the spectrum from $E_{\nu}^{\rm min}=0.8\,m_{\chi}$ to $E_{\nu}^{\rm max}=1.2\,m_{\chi}$, corresponding to twice the detector energy resolution around the line signal, while for the $\tau^{+}\tau^{-}$ channel we adopt $E_{\nu}^{\rm min}=0.33\,m_{\chi}$ to $E_{\nu}^{\rm max}=1.2\,m_{\chi}$, which encompasses the bulk of the DM signal while limiting contamination from the atmospheric background.

To set limits on the SD DM-neutron cross section, we assume Poisson-distributed events and impose a 95\% confidence level (CL) exclusion criterion. A given parameter point is excluded if the predicted total event count $N_{\chi}+N_{\rm bkg}$ is large enough that a background-only fluctuation producing this many events or more would occur with probability below 5\%, i.e.,
\begin{equation}
    \sum_{k=N_{\chi}+N_{\rm bkg}}^{\infty}\frac{\lambda^{k}e^{-\lambda}}{\Gamma(k+1)}\leq 0.05\,,
\end{equation}
where $\lambda \equiv N_{\rm bkg}(m_\chi)$ is the expected number of fiducial atmospheric muon-neutrino background events within the corresponding energy window. To avoid spurious limits in regions of parameter space where both the predicted signal and background are very small, we additionally require $N_{\chi}+N_{\rm bkg}\geq 2.7$ for an exclusion with 95\%~CL to be claimed~\cite{Feldman:1997qc}. Notably, our expected limits can be improved by the neutrino experimental collaborations, with dedicated analyses of both the DM annihilation signals and the astrophysical backgrounds from the Sun.

\begin{figure*}[tb]
    \centering
    \includegraphics[width=1\linewidth]{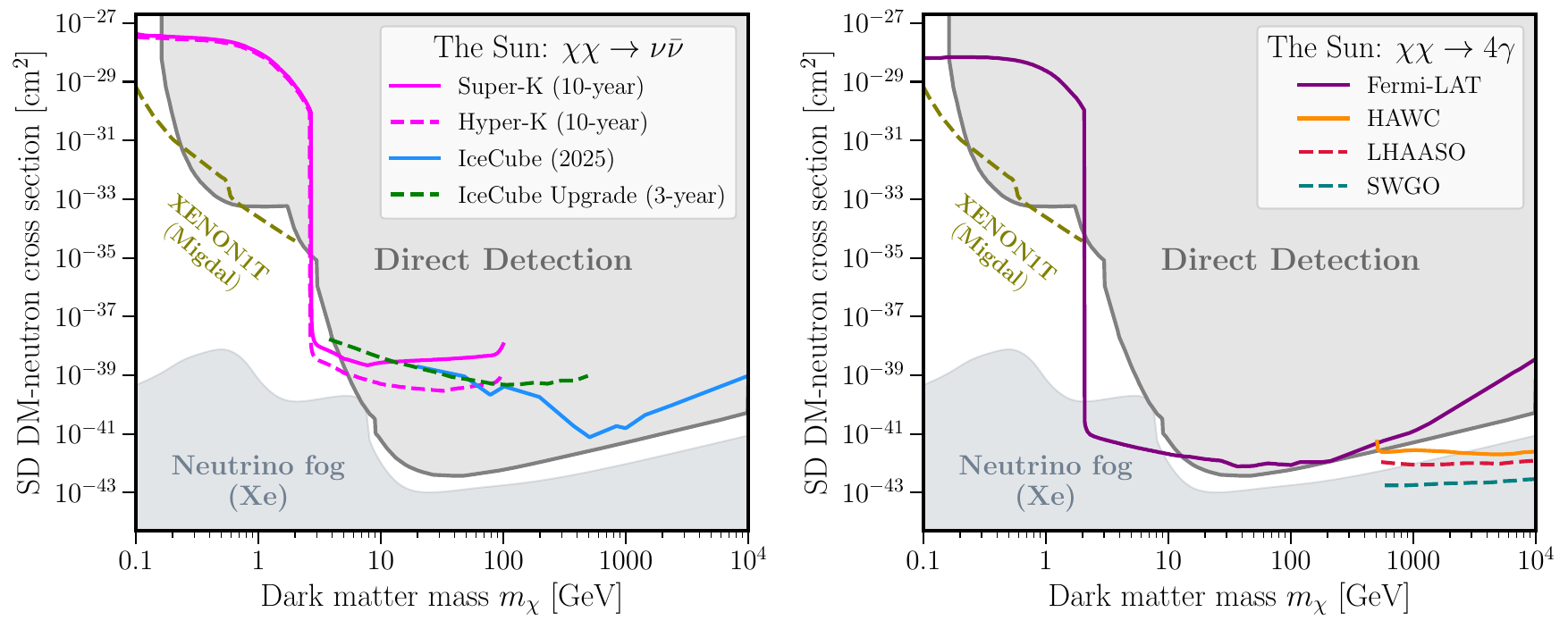}
    \caption{{\bf Left:} Neutrino constraints for the $\chi\chi\to \nu\bar{\nu}$ annihilation channel. Super-K constraints (solid) and Hyper-K projections with a 10-year solar observation are in magenta. Constraints using the IceCube 2025 upper limits are in blue, while IceCube Upgrade projections using a 3-year observation time are in green. {\bf Right:} $\gamma$-ray constraints for annihilation to long-lived particles that decay to the $2\gamma$ final state. Fermi-LAT constraints are in purple, HAWC constraints are in orange, while LHAASO projections are in red, and SWGO projections are the dashed teal line. The combined best limits from direct detection experiments are in gray, while the theoretical XENON1T Migdal-effect constraint is the dashed olive line. The xenon neutrino fog is the gray-blue shaded region. All solar constraints assume an $s$-wave annihilation cross section of $\langle\sigma v\rangle_{\chi\chi}=3\times10^{-26}$~cm$^{3}$/s, and the $\gamma$-ray constraints additionally assume unit branching into long-lived mediators with $P_{\rm surv}^{\gamma}=1$.}
    \label{fig:constraint_massless}
\end{figure*}

\subsubsection{IceCube and IceCube Upgrade}
\label{sssect:IceCube}

The IceCube Neutrino Observatory, located at the geographic South Pole, instruments a cubic kilometer of Antarctic ice with over 5000 digital optical modules deployed at depths between 1.45 and 2.45~km~\cite{IceCube:2016zyt}. Neutrinos are detected through the Cherenkov light of secondary charged particles, with muon tracks providing degree-scale angular resolution at TeV energies. The large instrumented volume makes IceCube the most sensitive detector for high-energy neutrinos, complementing the low-threshold reach of water Cherenkov detectors. On the other hand, its energy threshold of $\sim 100$~GeV (lowered to $\sim 10$~GeV by the denser DeepCore infill array~\cite{IceCube:2011ucd}) restricts its solar DM sensitivity to heavier DM masses. The IceCube Upgrade, currently being deployed within the DeepCore region, adds seven densely instrumented strings that will further lower the energy threshold to a few GeV and improve the reconstruction of GeV-scale events~\cite{Ishihara:2019aao}.

The IceCube Collaboration has been actively involved in the search for DM interactions in the Sun, in particular investigating the DM-proton and DM-electron scattering scenarios~\cite{IceCube:2021xzo, IceCube:2025fcu}. For several annihilation channels, the collaboration provides upper limits on the solar DM annihilation rate $\Gamma$ for DM masses above 10~GeV~\cite{IceCube:2025fcu}, with corresponding projected sensitivities for the IceCube~Upgrade extending down to 3.7~GeV~\cite{Hooper:2025ohk}, assuming a 3-year observation time~\cite{IceCube:2026rbh}. Notably, both mass ranges lie safely above the solar evaporation mass~\cite{Garani:2021feo}, so that the captured DM population is unaffected by evaporation across the entire range probed by these detectors.

To derive constraints on the DM-neutron cross-section, we begin by taking the current IceCube limits on the mass-dependent DM annihilation rate in the Sun (Table IV of~Ref.~\cite{IceCube:2021xzo}). We then utilize our capture, annihilation and evaporation calculations to convert these into limits on the SD DM-neutron scattering cross-section. Since DM annihilation produces similar fluxes in all neutrino flavors in our approximation for the $\nu\bar{\nu}$ channel, we combine the best limits among the flavor-specific results reported by IceCube. For IceCube, we use the limits derived in Ref.~\cite{IceCube:2025fcu}, while for the IceCube Upgrade, we adopt the projected limits from Ref.~\cite{IceCube:2026rbh}.

\subsection{Gamma-Ray Observations}
\label{ssect:gammaray_exp}

Solar $\gamma$-ray observations have emerged as a distinct branch of astroparticle physics over the past decade~\cite{Nisa:2019mpb}. The Sun shines in GeV--TeV $\gamma$-rays through two components: (1) disk emission from hadronic cosmic rays interacting with the solar atmosphere and photosphere~\cite{Zhou:2016ljf, Linden:2020lvz}, whose intensity, spectrum, as well as time- and angular-dependence remain poorly understood~\cite{HAWC:2022khj, Tang:2018wqp, Linden:2018exo, Cholis:2015gna}, and (2) an extended halo of $\gamma$-rays from the inverse-Compton emission from cosmic-ray electrons scattering off solar photons~\cite{Moskalenko:2006ta, Orlando:2008uk, Cholis:2020tpi, Linden:2025xom}, which is well-fit by theoretical models. Beyond these astrophysical components, solar $\gamma$-rays also provide a powerful probe of DM annihilation into long-lived mediators that decay outside the Sun~\cite{Leane:2017vag}. Here, we consider the space-based Fermi-LAT and the ground-based observatories HAWC and LHAASO, together with projections for the upcoming SWGO. We consider the cases where captured DM annihilates into long-lived particles that escape the Sun and then decay into two photons ($\chi\chi\to 4\gamma$) or an $e^{+}e^{-}$ pair ($\chi\chi\to 4e$).

\subsubsection{Fermi-LAT}
\label{sssect:fermi}

The Large Area Telescope on board the Fermi Gamma-ray Space Telescope (Fermi-LAT) is a pair-conversion detector that has continuously surveyed the $\gamma$-ray sky in the $\sim$0.1--1000~GeV range since 2008~\cite{Fermi-LAT:2009ihh}. Its wide field of view and all-sky coverage allow it to monitor the Sun throughout its orbit, making it the primary instrument for solar $\gamma$-ray observations at GeV energies. Analyses of more than a decade of Fermi-LAT data have measured the solar-disk emission in the 0.1--100~GeV range, finding an energy flux at the level of $\sim 2\times 10^{-8}$~GeV~cm$^{-2}$~s$^{-1}$~\cite{Linden:2020lvz}. This emission is thought to originate predominantly from hadronic cosmic rays interacting with the solar surface, and its observed time- and angular-dependent variations preclude a DM-dominated origin~\cite{Linden:2020lvz}.

We derive constraints using the upper limits from the Fermi-LAT solar-disk observation over the 0.1--100~GeV photon energy range, conservatively requiring the predicted $\gamma$-ray flux from DM annihilation into long-lived mediators not to exceed the observed flux. Since both decay channels considered here, $\phi\to 2\gamma$ and $\phi\to e^{+}e^{-}$, produce continuous spectra extending down to low photon energies, a portion of the signal always falls within the Fermi-LAT energy window even for DM masses well above 100~GeV. This allows us to extend the Fermi-LAT constraints up to DM masses of 10~TeV.

\subsubsection{HAWC, LHAASO, and SWGO}
\label{sssect:VHEgamma}

The High-Altitude Water Cherenkov (HAWC) observatory, located at an altitude of 4100~m on the Sierra Negra volcano in Mexico, consists of 300 water Cherenkov tanks covering an area of $\sim$22{,}000~m$^{2}$~\cite{historical:2023opo}. By detecting the shower particles that reach the ground, HAWC operates continuously with a wide field of view, enabling daytime observations of the Sun that are impossible for imaging atmospheric Cherenkov telescopes. Using its first years of data, HAWC derived dedicated constraints on DM annihilation into long-lived mediators from the solar direction~\cite{HAWC:2018szf}. Subsequently, HAWC successfully observed the solar disk at TeV energies during the brighter solar minimum period, finding astrophysical $\gamma$-ray emission from the Sun that extended up to energies of at least $\sim$3~TeV~\cite{HAWC:2022khj}.

The Large High Altitude Air Shower Observatory (LHAASO), located at an altitude of 4410~m in Sichuan, China, combines a $\sim$1~km$^{2}$ particle-detector array, a 78{,}000~m$^{2}$ water Cherenkov detector, and wide-field air Cherenkov telescopes, providing sensitivity to $\gamma$-rays from $\sim$100~GeV up to PeV energies~\cite{LHAASO:2019qtb}. Its ground-array technique likewise permits continuous observations of the solar direction, and its larger effective area is expected to extend the reach of solar $\gamma$-ray searches beyond that of HAWC at the highest energies. Although LHAASO has been operating for roughly five years, no solar-disk observation has yet been published by the collaboration. Since LHAASO is in principle more sensitive to this emission than HAWC, we use its projected sensitivity from Ref.~\cite{Leane:2017vag} to derive the projected limits presented in this work.

Finally, the Southern Wide-field Gamma-ray Observatory (SWGO) is a next-generation Cherenkov array with construction planned at Pampa la Bola in the Chilean Andes, at an altitude above 4700~m~\cite{Albert:2019afb}. As the first wide-field TeV observatory in the Southern Hemisphere, SWGO will complement HAWC and LHAASO with improved sensitivity, and its projected capabilities for solar observations have recently been studied in Ref.~\cite{Andrade:2024ekx}.

We use the limits from these ground-based $\gamma$-ray observatories to investigate DM masses from 500~GeV up to 10~TeV. We derive the HAWC constraints, as well as LHAASO and SWGO projections, by finding the cross-section values that overproduce the $\gamma$-ray fluxes from DM annihilation into long-lived particles inside the Sun, considering both decay channels of these particles, \mbox{$\phi\to 2\gamma$} and \mbox{$\phi\to e^{+}e^{-}$}.

\section{Constraints on Spin-Dependent Dark Matter-Neutron Cross Sections}
\label{sect:constraints}

\begin{figure*}[tb]
    \centering
    \includegraphics[width=1\linewidth]{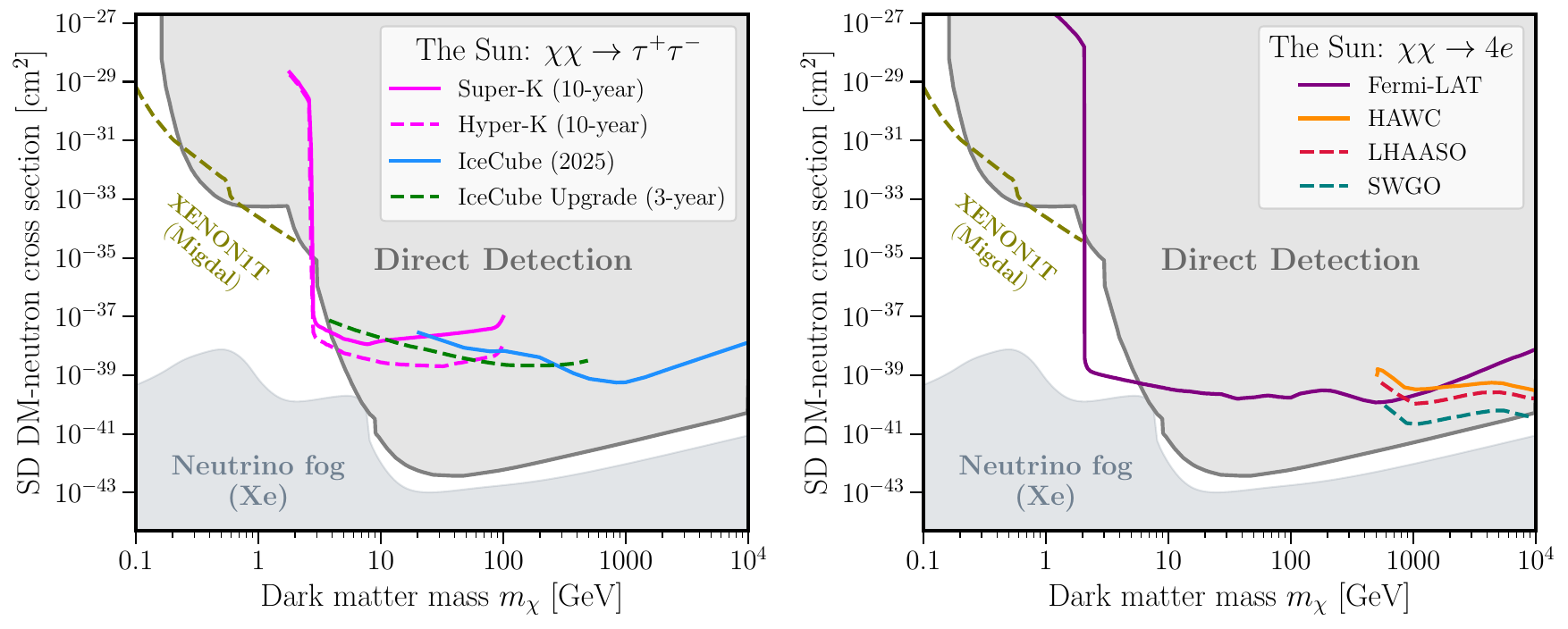}
    \caption{Similar to Fig.~\ref{fig:constraint_massless}, but for leptonic final states. {\bf Left:} Constraints for the $\chi\chi\to \tau^{+}\tau^{-}$ annihilation channel with neutrino observations. {\bf Right:} Constraints for the case where long-lived mediators from DM annihilation decay to $e^{+}e^{-}$ with $\gamma$-ray observations. The thermal $s$-wave annihilation cross section is assumed to be \mbox{$\langle\sigma v\rangle_{\chi\chi}=3\times 10^{-26}$~cm$^{3}$/s}, and the survival probability of the long-lived mediator is $P_{\rm surv}^{\gamma}=1$.}
    \label{fig:constraint_lepton}
\end{figure*}

Figures~\ref{fig:constraint_massless} and \ref{fig:constraint_lepton} show our constraints on the SD DM-neutron cross section from solar observations with both neutrino and $\gamma$-ray experiments for all annihilation channels considered in this work. We compare these solar constraints with the combined best limits from direct detection experiments: LUX-ZEPLIN (LZ)~\cite{LZ:2024zvo}, XENONnT~\cite{XENON:2025vwd}, PandaX~\cite{PandaX:2024qfu}, CDMSlite~\cite{SuperCDMS:2018gro}, and CRESST~\cite{CRESST:2022dtl}. We also show XENON1T limits utilizing the Migdal effect~\cite{XENON:2019zpr}. Finally, we indicate the neutrino fog, computed for SD DM-neutron couplings on a xenon target following Ref.~\cite{OHare:2021utq}, below which solar neutrino backgrounds significantly hamper terrestrial direct detection searches. We note that our solar constraints are computed with $\rho_{\chi}=0.4$~GeV/cm$^{3}$ and the halo parameters of Sec.~\ref{ssect:capt_eva}, whereas the direct detection limits shown are published assuming $\rho_{\chi}=0.3$~GeV/cm$^{3}$ and the SHM++ parameters of Ref.~\cite{Evans:2018bqy}; rescaling our constraints to the direct detection conventions would weaken them at the tens-of-percent level, without affecting our conclusions.

Figure~\ref{fig:constraint_massless} (left panel) shows our solar constraints for the $\chi\chi\to \nu\bar{\nu}$ annihilation channel using neutrino experiments. Current IceCube solar observations produce conservative limits from 10~GeV up to 10~TeV, while the projected sensitivity of the IceCube Upgrade with a 3-year observation extends our sensitivity down to DM masses of 3.7~GeV. Both of these constraints, however, lie within the parameter space already excluded by current direct detection. At lighter masses, Super-K constrains new cross sections down to \mbox{$\sim 10^{-38}$~cm$^{2}$} for DM masses in the range 2.5--5~GeV, and future Hyper-K observations can improve on these limits by roughly a factor of three. 

Notably, these low-threshold neutrino detectors extend the limits down to 100~MeV, covering the mass range from 100--200~MeV where direct detection loses sensitivity. The steep weakening of solar constraint curves below $\sim$2.5~GeV comes from the solar evaporation mass discussed in Sec.~\ref{ssect:capt_eva}. Below this mass, evaporation depletes the largely isothermal DM population, and only DM in local thermal equilibrium with large annihilation fluxes remains observable. The XENON1T limits based on searches for a signal induced by the atomic Migdal effect extend down to 85~MeV. The atomic Migdal effect was long predicted theoretically and has recently been observed in neutron scattering on a gaseous low-Z target~\cite{Yi:2026fmf}. However, the Migdal response in liquid xenon itself has not yet been directly measured. The solar constraints therefore provide an independent astrophysical probe of this mass window. This search strategy, combined with lower-threshold neutrino experiments, could extend the solar constraints to even lighter DM masses~\cite{Bernal:2012qh, Rott:2012qb}.

Figure~\ref{fig:constraint_massless} (right panel) shows $\gamma$-ray constraints for the \mbox{$\chi\chi\to2\phi\to4\gamma$} annihilation channel. The Fermi-LAT constraints span the widest mass range, from 100~MeV up to 10~TeV, owing to the broad box-shaped spectrum of the mediator decay. These constraints surpass current direct detection limits in the 2--12~GeV range and reach down to $\sim10^{-42}$~cm$^{2}$. Notably, in this mass range the Fermi-LAT constraints reach into the xenon neutrino fog, probing parameter space where terrestrial direct detection searches face an irreducible solar neutrino background. At the highest masses, the ground-based observatories take over. HAWC and LHAASO constrain the cross section from 500~GeV up to 10~TeV and surpass direct detection limits, while the projected SWGO sensitivity improves on these constraints and reaches down to $\sim 10^{-43}$~cm$^{2}$ for TeV DM masses. Similar to the neutrino case, the $\gamma$-ray constraints extend down to 100~MeV, and future MeV $\gamma$-ray missions such as COSI~\cite{Tomsick:2019wvo} and the proposed AMEGO~\cite{Caputo:2017sjw} and e-ASTROGAM~\cite{e-ASTROGAM:2017pxr} could extend the solar constraints to even lighter DM masses.

Finally, Fig.~\ref{fig:constraint_lepton} shows similar constraints on the SD DM-neutron cross section, but for DM annihilation to leptonic final states. Since neutrinos and $\gamma$-rays are only secondary products of these annihilation channels, the expected indirect detection signals are suppressed compared to the previous cases. This is reflected in the heavy DM mass regime, where the HAWC, LHAASO, and SWGO results are suppressed and already excluded by direct detection limits. However, at lighter masses, from 2.5--4~GeV for neutrinos and 2--5~GeV for $\gamma$-rays, current observations from Super-K and Fermi-LAT, along with future Hyper-K observations, remain sensitive to a new cross section regime, indicating the advantage of solar observations in probing this SD DM-neutron interaction.

\section{Conclusion}
\label{sect:conclusion}

In this paper, we provide a detailed study of the SD DM-neutron interaction in the Sun. For the first time, we compute the solar capture and evaporation rates for this coupling, which plays a central role in terrestrial direct detection but has so far been neglected in the solar DM capture literature. Using the AGSS09 standard solar model, we identify the dominant odd-neutron targets, $^{3}$He, $^{13}$C, $^{17}$O, $^{21}$Ne, $^{25}$Mg, and $^{29}$Si, and compute the capture and evaporation rates with a self-consistent treatment of the captured DM distribution for an arbitrary SD DM-neutron cross section.

We find that the capture rate is dominated by $^{3}$He for DM masses below $\sim 40$~GeV and is suppressed by 4--5 orders of magnitude relative to the SI and SD DM-proton and DM-electron cases, directly reflecting the trace abundances of the odd-neutron isotopes. As a consequence, the transition cross section at which the Sun becomes optically thick is pushed from the canonical value of $\sim 10^{-35}$~cm$^{2}$ up to $\sim 10^{-31}$~cm$^{2}$, a distinctive feature of this interaction. Despite this suppression, the evaporation mass remains around 2--4~GeV, similar to the proton scattering case, leaving a wide mass range where the captured DM population survives to annihilate.

Confronting the predicted annihilation signals with current neutrino and $\gamma$-ray solar observations, we derive the first observational constraints on the SD DM-neutron cross section from a celestial body. For the $\chi\chi\to\nu\bar{\nu}$ channel, Super-K opens new parameter space beyond direct detection in the 2.5--5~GeV mass range, reaching cross sections down to $\sim 10^{-38}$~cm$^{2}$, while future Hyper-K data can improve these limits by a factor of three. For the $\chi\chi\to 2\phi\to 4\gamma$ channel, the Fermi-LAT constraints surpass direct detection limits in the 2--12~GeV range, reaching down to $\sim 10^{-42}$~cm$^{2}$ and extending into the xenon neutrino fog, while HAWC, LHAASO, and upcoming SWGO data extend the coverage up to 10~TeV. Both messengers also probe the 100--200~MeV mass range where direct detection loses sensitivity. Even for the suppressed leptonic channels, solar observations remain sensitive to new parameter spaces at low DM masses.

Our results demonstrate that the SD DM-neutron interaction remains firmly within the reach of solar observations. Other current and future low-threshold neutrino experiments, such as JUNO~\cite{JUNO:2025fpc}, Borexino~\cite{BOREXINO:2023ygs}, KamLAND~\cite{KamLAND:2021gvi}, SNO+~\cite{SNO:2021xpa}, Daya Bay~\cite{Band:2013zka}, and DUNE~\cite{Capozzi:2018dat}, as well as MeV $\gamma$-ray missions such as COSI~\cite{Tomsick:2019wvo} and e-ASTROGAM/AMEGO~\cite{e-ASTROGAM:2017pxr, Caputo:2017sjw}, could extend these constraints to even lighter DM masses, while dedicated solar analyses by the experimental collaborations could substantially strengthen the conservative limits presented here. At the high energy frontier, our results can also motivate solar DM searches with other experiments and proposals such as ANTARES~\cite{ANTARES:2016obx}, KM3NeT~\cite{KM3NeT:2024xca}, P-ONE~\cite{Malecki:2024tvt}, TRIDENT~\cite{Wang:2026kka}, and IceCube-Gen2~\cite{IceCube-Gen2:2023vtj}.

Finally, while we choose the Sun to study the SD DM-neutron interaction in this work, this interaction can be further probed with other neutron-rich objects, such as neutron stars~\cite{Leane:2021ihh, Acevedo:2024ttq} and white dwarfs~\cite{Acevedo:2023xnu}. With their lower evaporation masses, these objects can target the unexplored window of lighter DM below 2~GeV~\cite{Garani:2021feo}. Overall, the rich opportunities offered by current and future observations of different celestial objects prove that the SD DM-neutron interaction can be thoroughly probed, and is far from a lost case.

\begin{acknowledgments}
We thank Lillian Santos-Olmsted and Tim Tait for fruitful discussions. TTQN especially thanks the TASI organizers and the University of Colorado Boulder for their hospitality during the period in which a large portion of this work was completed. TTQN and TL are supported by the Swedish Research Council under contract 2022-04283. TTQN is also supported by two grants from the Royal Swedish Academy of Sciences (KVA): PH2025-0073 (Physics) and AST2025-0048 (Astronomy and Space Science). TL is also supported by the Swedish National Space Agency under contract 117/19. TTQN and TL acknowledge the support from EDUCATE Excellence Centre funded by the Swedish Research Council through grant Dnr~2022-06627.\\

\noindent {\bf AI Usage Statement.} Claude Code was used to assist with generating the plots. Claude was used for grammar and typos checks. AI tools were not used to generate scientific results or draw conclusions. All analysis, code, and the final text were produced, checked, and verified by the authors, who take the full responsibility for the content of this work.
\end{acknowledgments}

\bibliography{ref}
\end{document}